\documentclass[12pt]{elsarticle}
\begin{document}

\thispagestyle{empty}
\begin{center}
{\Large\bf{Transverse double-spin asymmetries \\
\vskip 0.2cm
for electroweak gauge-boson  production \\
\vskip 0.2cm
in high-energy polarized $p^{\uparrow} + p^{\uparrow}$ collisions}}

\vskip 1.0cm
{\bf Jacques Soffer and Bernd Surrow}
\vskip 0.2cm
Temple University, Department of Physics, 1925 N. 12th St., Philadelphia, 
PA 19122, USA\\

\vskip 0.4cm
{\bf Claude Bourrely}
\vskip 0.2cm
Aix Marseille Univ, Univ Toulon, CNRS, CPT, Marseille, France\\
\vskip 0.5cm

{\bf Abstract}
\end{center}
We consider the production of $W$ and $Z / \gamma^{*}$ gauge bosons in proton-proton collisions at a center-of-mass 
energy of $\sqrt{s} = 510 \mbox{GeV}$ available at RHIC at BNL, operating at high luminosity. 
We stress the importance of measuring the transverse double-spin asymmetries $A_{TT}$, in connection 
with available transversely polarized beams with a high degree of polarisation. We will 
discuss some theoretical issues related to the predicted asymmetries. These studies are contrasted 
to the 2017 RHIC running operation of transversely polarized beams of mid-rapidity $W$ and $Z$ 
boson production at a center-of-mass energy of $\sqrt{s}=510\,$GeV along with long-term prospects beyond 2020 at RHIC.


\newpage 
\section{Introduction}
The collision of high-energy polarized proton beams with a
center-of-mass energy of $\sqrt{s}=200\,$GeV and $\sqrt{s}=510\,$GeV 
at the Relativistic Heavy-Ion Collider (RHIC) at Brookhaven National
Laboratory (BNL) provides a powerful way to gain a deeper insight into the spin structure
and dynamics of the proton such as the study of the helicity distributions of gluons
and quarks/antiquarks based on well established high-energy QCD and 
$W$ boson processes, respectively \cite{Adamczyk:2014xyw, Adamczyk:2014ozi, Adare:2014hsq, Adare:2015gsd}. Several studies have been suggested in the past to 
gain a better understanding of the quark transversity distribution, in particular the 
measurement of the transverse double-spin asymmetries ($A_{TT}$) for $W$ and $Z$ 
production \cite{Bourrely:1994sc}. At lowest-order (LO) it was found that $A_{TT}=0$ 
for $W$ production and the estimate of $A_{TT}$ for $Z$ production was done by assuming 
that the transversity distributions $h_1^q (x)$ coincide with the helicity distributions $\Delta q(x)$, similarly for antiquarks. 
This simplification is valid for a proton at rest, in the 
absence of relativistic effects, but in general these distributions are different. 
Among the three fundamental distributions of twist-2 are the unpolarized quark distributions 
$q(x)$, the quark helicity distributions $\Delta q(x)$ and the quark transversity distributions
$h_1^q (x)$. Transversity distributions are the least known because it suffers from the lack of experimental
information. The distributions $h_1^q (x)$ are chiral odd and therefore they are not directly accessible 
in Deep Inelastic Scattering (DIS). Since there is no gluon transversity distribution in a 
nucleon, the quark transversity distribution has a pure non-singlet $Q^2$ evolution.\\
Several years ago \cite{Soffer:1994ww}, the following positivity bound
\begin{equation}\label{Delta}
q (x) + \Delta q (x) \ge 2 | h_1^q (x) |,
\end{equation}
(similarly for antiquarks), was obtained at LO. 
Later \cite{Bourrely:1997bx, Vogelsang:1997ak, Martin:1997rz}, 
this strong result was proven to be valid after $Q^2$ evolution, up to next-to-leading order (NLO). \\
The transverse double-spin asymmetry is defined as $A_{TT}=(d\sigma_{\uparrow\uparrow}-
d\sigma_{\uparrow\downarrow})/(d\sigma_{\uparrow\uparrow}+d\sigma_{\uparrow\downarrow})$, 
where $\sigma_{\uparrow\uparrow} (\sigma_{\uparrow\downarrow})$ denotes the cross section with
the two initial protons transversely polarized in the same
(opposite) direction. This physics observable was considered earlier for high-$p_T$ jet and direct 
photon production and was estimated at RHIC energy at the $10^{-3}$ level \cite{Jaffe:1996ik}. 
The same order of magnitude was obtained later for upper bounds on $A_{TT}$, by saturating the 
transversity distributions, at low scale by the positivity bound Eq. (\ref{Delta}) \cite{Soffer:2002tf}. The 
experimental finding of large asymmetries would constitute one more spin puzzle and it is also  
very challenging in the case of gauge boson production.
In view of the 2017 running operation at RHIC of transversely polarized proton beams 
and the fact that the reconstruction of $W$ and $Z$ bosons is well established at mid-rapidity 
by the STAR collaboration \cite{STAR:2011aa}, we thought it is timely to revisit this subject.\\ 
In this letter, we will extend the discussion beyond the LO case for the prediction of $A_{TT}=0$ for $W$ production.
The main purpose is to present a NLO calculation of the allowed maximal $A_{TT}$ for 
$Z$ production, using Eq. (\ref{Delta}). We will also give some experimental details concerning the
future measurement of these asymmetries, as well as the expected accuracy.

\section{Transverse double-spin asymmetry for gauge boson production and quark transversity distributions}

For gauge boson production, assuming that the underlying parton
subprocess is quark-antiquark annihilation, we easily find at LO
\begin{equation}
A_{TT}=\hat{a}_{TT}\times\frac{\sum_{i=u,d}\left(b_i^2-a_i^2\right)
\left[ h_1^{q_i}(x_1) h_1^{\bar q_i}(x_2)+(x_1\leftrightarrow x_2)\right]}
{\sum_{i=u,d}\left(a_i^2+b_i^2\right)
\left[ q_i(x_1)\bar q_i(x_2)+(x_1\leftrightarrow x_2)\right]}.
\label{ATT}
\end{equation}
Here $\hat{a}_{TT}$ is the partonic transverse double-spin asymmetry, given in Ref. \cite{Ji:1992ev}, which reads
\begin{equation}
\hat{a}_{TT} = \frac{\sin^{2}{\theta_{cm}} \cos{2 \phi}_{cm}}{ 1 + \cos^{2}{\theta_{cm}}},
\label{aTT}
\end{equation}
a simple expression in terms of $\theta_{cm}$ and $\phi_{cm}$, the polar and azimuthal angles of one charged lepton momentum
with respect to the beam and the proton polarization, respectively. Clearly $\hat{a}_{TT} = 1$, its maximum value, is obtained when
$\theta_{cm} = \pi/2$ and $\phi_{cm} = 0$. However in order to increase the observable events rate, one should integrate over the angles and one gets $\hat{a}_{TT} = 1/\pi$. In order to extend Eq. (\ref{ATT}) to NLO, one has to introduce the NLO transversity distributions  $h_{1}^{q}(x,Q^2)$ and to replace $\hat{a}_{TT}$, by its NLO expression $\hat{a}_{TT}({\cal{O}}(\alpha_{s}))$, following the results of Ref. \cite{Martin:1999mg}.\\
In Eq. (\ref{ATT}) $a_i$ and $b_i$ denote the vector and axial-vector coupling constants of $q_i$ to the produced gauge boson.
This result generalizes the case of lepton-pair production, i.e. Drell-Yan production, 
through an off shell photon $\gamma^{\star}$ and corresponding
to $b_i=0$ and $a_i=e_i$, the electric charge of $q_i$. For
$W^{\pm}$ production, which is purely left-handed, we expect $A_{TT}=0$,
since in this case $a_i^2=b_i^2$. This strong result which is valid beyond LO (see Appendix) is worth checking
experimentally. It follows from the fact that one assumes a universal $V - A $ coupling. The observation of a non-zero $A_{TT}$, could be 
the footprints of new physics at high scales, beyond the Standard Model.\\
NLO corrections to the above expression Eq. (\ref{ATT}) have been calculated \cite{Martin:1997rz} and they gave 
estimates of the maximal $A_{TT}$ for Drell-Yan dimuon production of mass $M$. Upper bounds on $A_{TT}$ as function 
of the dimuon rapidity, for $M = 5 -20 \mbox{GeV}$, much below the $Z$ peak, were also obtained  \cite{Martin:1999mg}.\\
We will now summarize what is known about the transversity distributions, from two different methods of indirect extraction. Transversity can be extracted in the standard framework of collinear
factorization using SIDIS with two hadrons detected in the final state. In this case, the distribution $h_1^q (x)$ is
multiplied by another specific chiral-odd Di-hadron Fragmentation Function (DiFF), which can be extracted from the corresponding $e^{+}e^{-}$ annihilation process leading to two back-to-back hadron pairs. In the collinear framework, evolution equations of DiFFs can
be computed and in particular, in Ref. \cite{Radici:2015mwa} they
have considered the recent measurement from the COMPASS collaboration for identified
hadron pairs produced off transversely polarized proton targets  \cite{Braun:2015baa}. By combining proton and deuteron data, COMPASS has managed to perform the flavor separation $h_1^u (x)$ and $h_1^d (x)$. The error analysis is performed with the so-called replica method based on the random generation of a large number of replicas of the experimental points. As such, this method allows for a more realistic estimate of the uncertainty on DiFFs, so the results shown in  Ref. \cite{Radici:2015mwa} represent certainly the currently most realistic estimate of the uncertainties on the valence transversity distributions (see Fig.8 of \cite{Radici:2015mwa}).\\
\begin{figure}[t]
\begin{center}
\includegraphics[width=11.0cm]{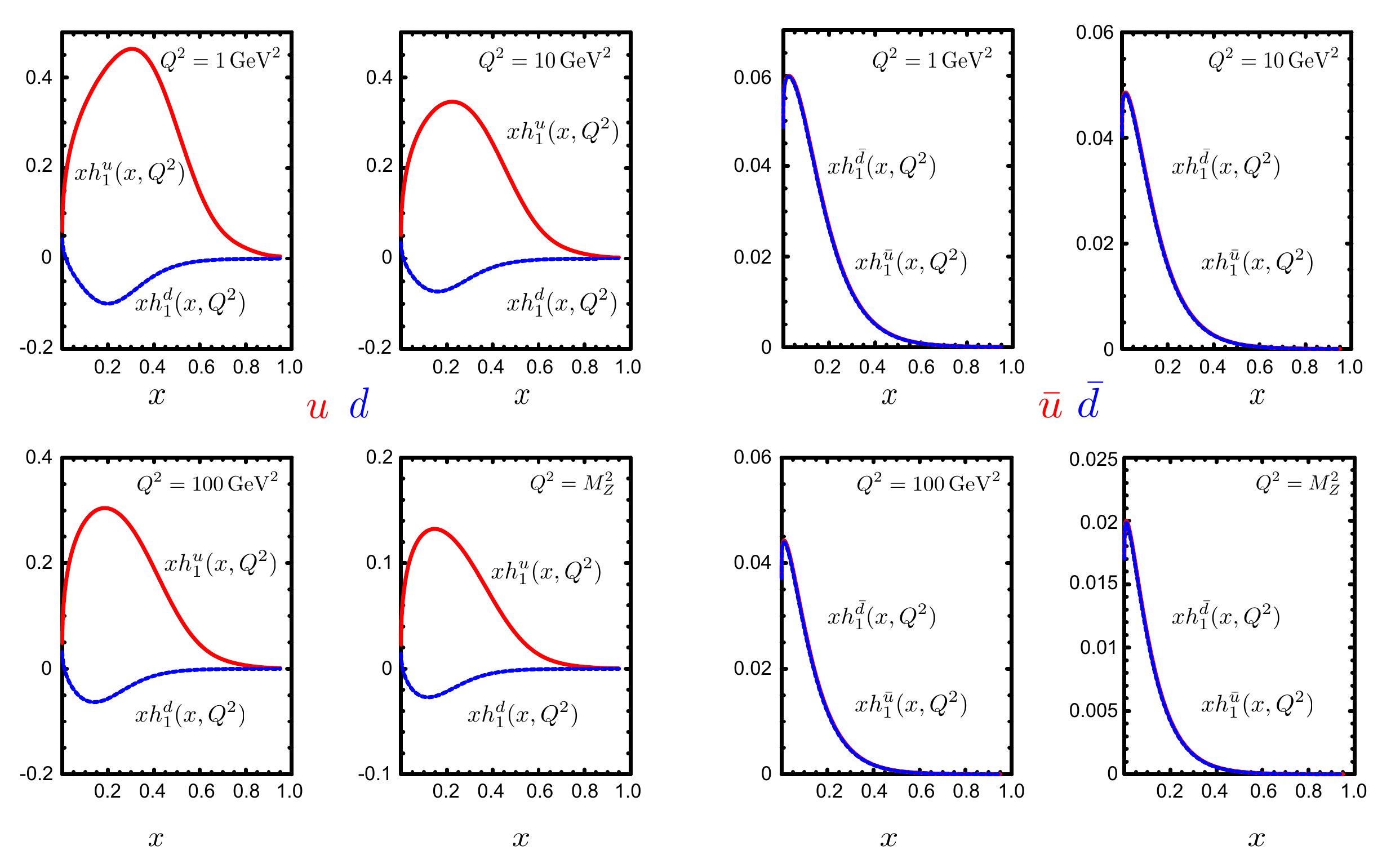}
\end{center}
\caption{\small  ({\it left}) Maximal bounds for quark transverse distributions $h_{1}^{q}(x,Q^2)$ $(q=u,d)$, from the PDFs set of BS15 \cite{Bourrely:2015kla}, as a function of
$x$, at the initial scale $Q^2=1 \mbox {GeV}^2$,  $Q^2=10 \mbox {GeV}^2$,  $Q^2=100 \mbox {GeV}^2$ and at $Q^2= M_Z^2$, after
QCD NLO evolution. ({\it right}) Same for antiquark transverse distributions. Note that they are equal for $\bar u$ and $\bar d$, which is a consequence
of an approximate equality of the potentials $X_u^- = X_d^-$ in the BS model, as found in Ref. \cite{Bourrely:2015kla} and earlier works.}
\label{fig1}
\end{figure}

Transversity can be also extracted using another method involving transverse momentum dependent (TMD) PDFs, in particular the Collins fragmentation function. Assuming TMD factorization, the Collins fragmentation function (FF) can be studied in SIDIS experiments,
where it appears to be convoluted with the transversity distribution $h_1^q (x,p_T)$ depending on $p_T$, which is now considered as a TMD PDF. It induces an asymmetry, the so-called Collins azimuthal asymmetry, which has been clearly observed experimentally. The Collins FF also induces azimuthal angular correlations between hadrons produced in opposite jets in $e^{+}e^{-}$  annihilation. Consequently, a simultaneous analysis of SIDIS and $e^{+}e^{-}$  data allows the combined extraction of the transversity distributions and the Collins FFs. The most recent results of this method can be found in Refs. \cite{Anselmino:2015sxa, Kang:2015msa}.\\
Their fit results are compatible (see Fig. 6 of Ref. \cite{Anselmino:2015sxa} and Fig. 3 of Ref. \cite{Kang:2015msa}) and one finds that  for $Q^2 =2.4 \mbox{GeV}^2$, $h_1^u (x)>0$ has a maximum of 0.3 for $x$ near 0.4 and $h_1^d (x)<0$ has a maximum of -0.15 for $x$ near 0.1. For the down quark the tranversity tends to saturate the positivity bound, for $x>0.2$.
We emphasize the fact that nothing is known for antiquarks transversity.\\
Our purpose is to present a NLO calculation of the allowed maximal $A_{TT}$ for 
$Z$ production, using Eq. (\ref{Delta}) to construct the maximal bound for quark (antiquark) transversity distributions $h_{1}^{q}(x,Q^2)$ $(q=u,d)$, from the PDFs set of BS15 \cite{Bourrely:2015kla}. In Fig. \ref {fig1} we display these distributions which saturate the positivity bound at the initial scale $Q^2 = 1 \mbox{GeV}^2$ and at $Q^2= M_Z^2$, after QCD NLO evolution. We observe a rather strong reduction at this high energy scale. They will be used to calculate the allowed maximal $A_{TT}$ for 
$Z$ production at RHIC at BNL and the results are given in the next section, together with the expected accuracy on this measurement.
Finally let us also mention recent lattice-QCD results \cite{Chen:2016utp} on the isovector transversity distribution $x[h_1^u (x) - h_1^d (x)]$
\cite{Chen:2016utp}, using the large-momentum effective field theory (LaMET) approach. It leads to a surprising large result in the large-$x$ region and to some insight into the corresponding antiquark quantity (see Fig. 6 of Ref. \cite{Chen:2016utp}).

\section{Results}
The production of $Z$ bosons in polarized proton collisions at $\sqrt{s}=500\,$GeV has been demonstrated in a proof-of-principle 
measurement by the STAR collaboration \cite{STAR:2011aa}. The measurement 
of the total cross-section is based on an integrated luminosity of $13.2\,$pb$^{-1}$ of a data sample collected in 2009 with an estimated 
total efficiency of $\epsilon^{tot}=0.338 \, \pm \, 0.012 \, \rm{(stat.)}\, \pm \, 0.021\, \rm{(sys.)}$. We denote by $\eta$ and $E_T$ the pseudorapidity and the transverse energy of a charged lepton in the $Z/\gamma^{*}\rightarrow e^{+}e^{-}$ production. The total number of observed events in the mid-rapidity region of $|\eta|<1$,
$E_{T}>15\,$GeV and an invariant $e^{+}e^{-}$ mass window cut of $70\,<m_{\rm inv}\,<\,110\,$GeV amounts to $13\pm 3.6$
with a small background contribution of $0.1\, \pm 0.1 \, \rm{(stat.)}\, ^{+1.3\,\rm{(sys.)}}_{-0.0\,\rm{(sys.)}}$. The fiducial cross-section is given by $\sigma_{Z / \gamma^{*}}^{fid}\cdot {\rm BR}(Z / \gamma^{*}\rightarrow e^{+}e^{-}) = 2.9\, \pm 0.8 \, \rm{(stat.)}\, ^{+0.2\,\rm{(sys.)}}_{-0.3\,\rm{(sys.)}}\,\pm 0.4 \, \rm{(lumi)}\,$pb which translates into a total cross-section of $\sigma_{Z / \gamma^{*}}^{tot}\cdot {\rm BR}(Z / \gamma^{*}\rightarrow e^{+}e^{-}) = 7.7\, \pm 2.1 \, \rm{(stat.)}\, ^{+0.5\,\rm{(sys.)}}_{-0.9\,\rm{(sys.)}}\,\pm 1.0 \, \rm{(lumi)}\,$pb taking into account acceptance corrections for fidicial and kinematic requirements. This measured cross-section agrees within uncertainties with 
a quoted theoretical predication of $10.8\,$pb at NLO level based on the unpolarized PDF set of MSTW08 \cite{STAR:2011aa}. A larger data 
sample taken in 2011 and 2012 of $86\,$pb$^{-1}$ was used to measure the first longitudinal single-spin asymmetry $A_{L}$ for $Z/\gamma^{*}\rightarrow e^{+}e^{-}$ 
production for $|\eta|<1$, $E_{T}>14\,$GeV and an invariant $e^{+}e^{-}$ mass window cut of $70\,<m_{\rm inv}\,<\,110\,$GeV \cite{Adamczyk:2014xyw}. The measured asymmetry of 
$A_{L}^{Z/\gamma^{*}}=-0.07^{+0.14}_{-0.14}$ is consistent within the large uncertainties with NLO calculations based on different helicity-dependent 
parton distribution functions.
\begin{figure}[t]
\begin{center}
\includegraphics[width=9.0cm]{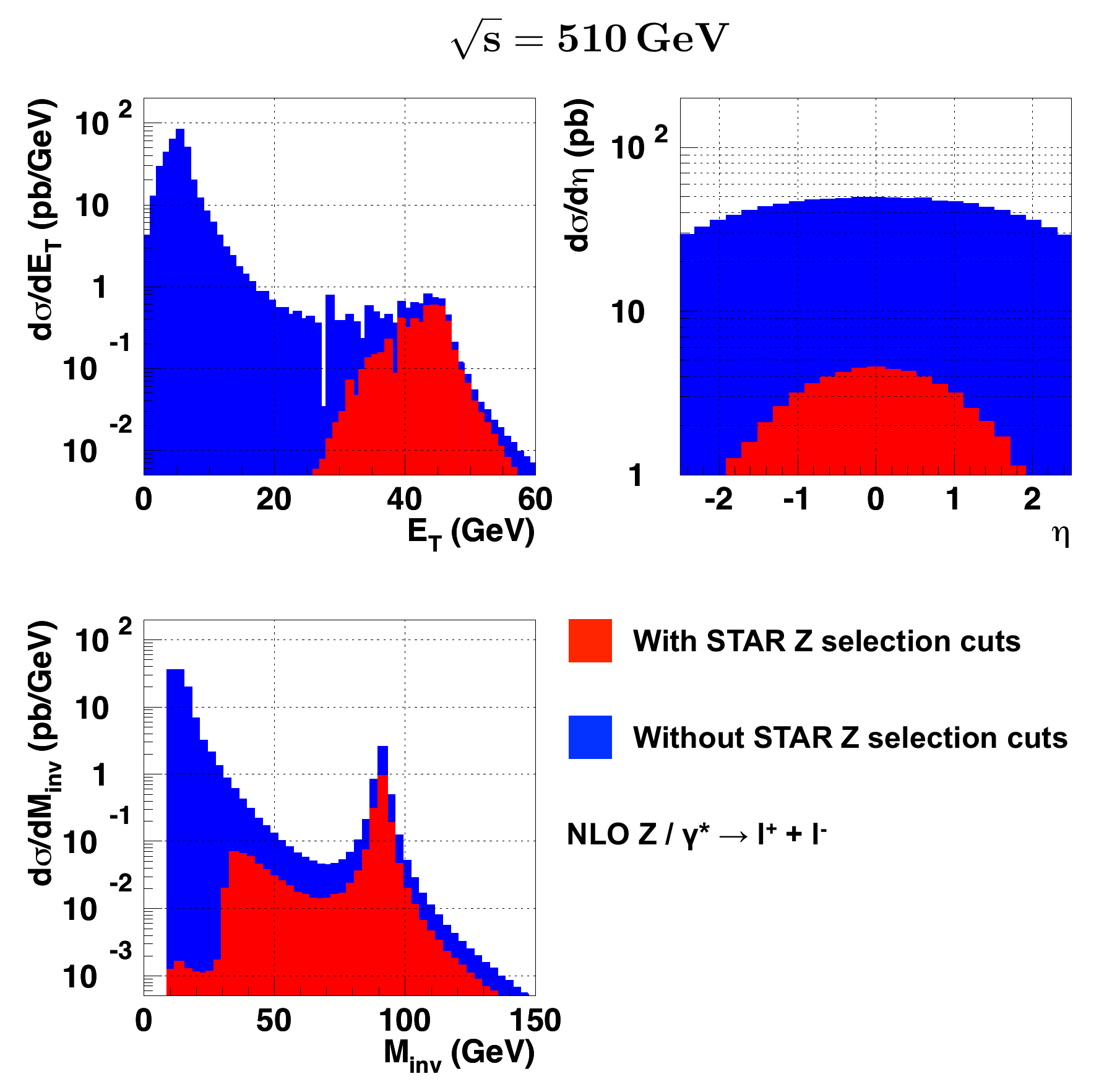}
\end{center}
\caption{\small Various NLO cross sections related to $Z$ production, blue shade with no cuts, red shade with specific cuts. The $E_T$ distribution (upper left corner) with  $70\,<M_{\rm inv}\,<\,110\,$GeV and $|\eta|<1$, the pseudo rapidity distribution (upper right corner) with  $70\,<M_{\rm inv}\,<\,110\,$GeV and $E_{T}>15\,$GeV, the invariant mass distribution (lower left corner) with $E_{T}>15\,$GeV and $|\eta|<1$.}
\label{fig2}
\end{figure}
\begin{figure}[t]
\begin{center}
\includegraphics[width=9.0cm]{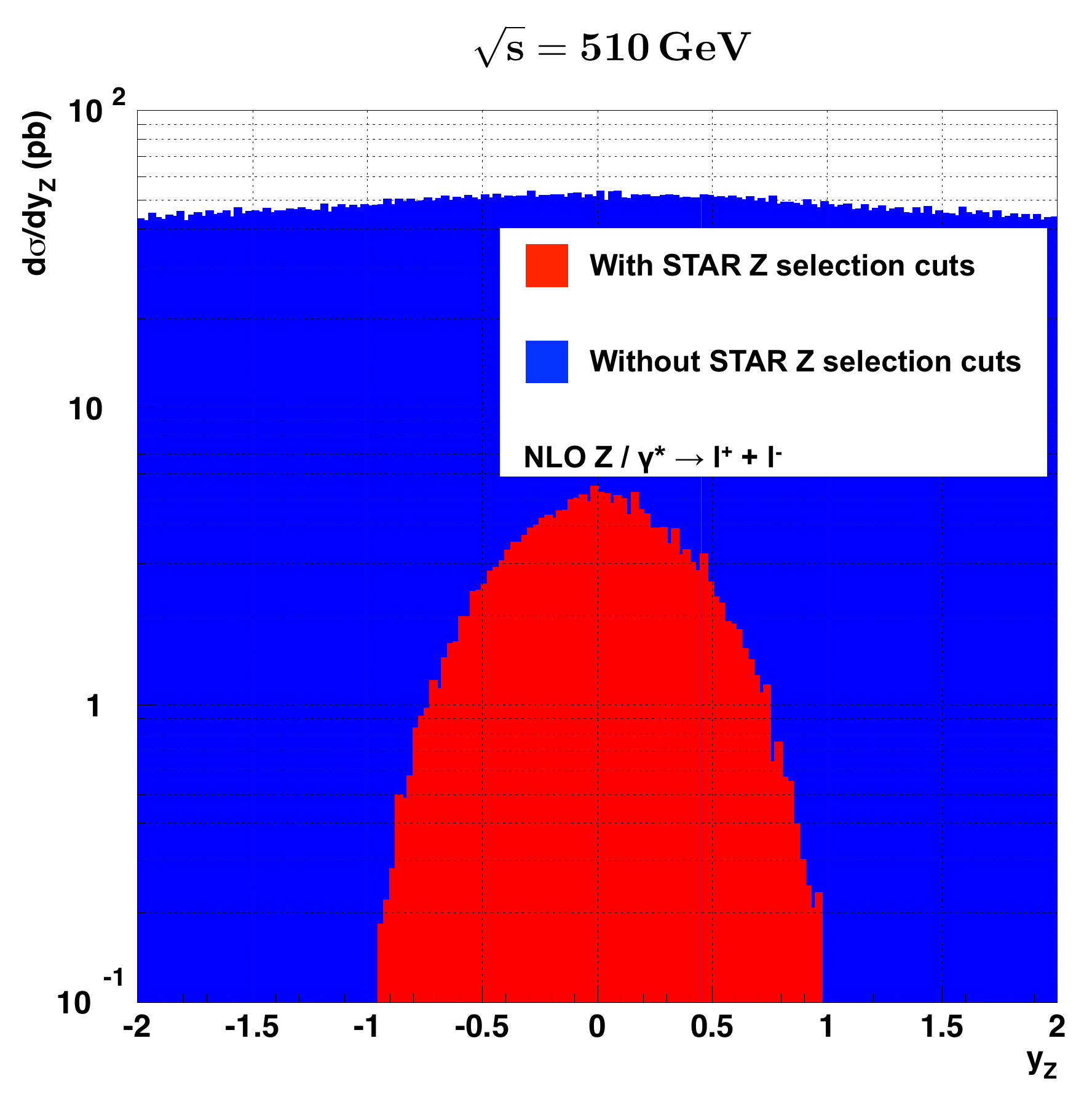}
\end{center}
\caption{\small  Differential NLO cross-section of $Z$ boson rapidity as a function of $y_{Z}$, blue shade with no cuts, red shade with specific cuts.}
\label{fig3}
\end{figure}

The measurement of both $W$ and $Z$ production is well established at mid-rapidity by the STAR experiment. The phase space cuts which will be used for determining
$A_{TT}$ for $Z/\gamma^{*}\rightarrow e^{+}e^{-}$ are based on the STAR selection cuts of $|\eta|<1$, $E_{T}>15\,$GeV and an 
invariant $e^{+}e^{-}$ mass window cut of $70\,<m_{\rm inv}\,<\,110\,$GeV. Figure~\ref{fig2} shows the $E_{T}$, $\eta$ and $m_{\rm inv}$
distributions for $Z/\gamma^{*}\rightarrow e^{+}e^{-}$ production with the characteristic Jacobian peak for $Z$ production at roughly half of the
$Z$ boson mass in the $E_{T}$ distribution and a peak at the mass of the $Z$ boson in the $m_{\rm inv}$ distribution. 
These NLO distributions have been determined using a NLO MC framework
by D.~de Florian and W. Vogelsang \cite{deFlorian:2010aa} allowing the calculation of various differential distributions at NLO level. The Z boson rapidity distributions with and without STAR selection cuts are shown in Figure~\ref{fig3}.  Note that we have the symmetry property $d\sigma/dy_{Z}(y_{Z})=d\sigma/dy_{Z}(-y_{Z})$. Extensions of the measurable $\eta$ coverage for $e^{+}/e^{-}$ would be
advantages to increase the $y_{Z}=(1/2)(\eta_{e^{+}}+\eta_{e^{-}})$ coverage.

\begin{figure}[t]
\vspace*{-2cm}
\begin{center}
\includegraphics[width=9.0cm]{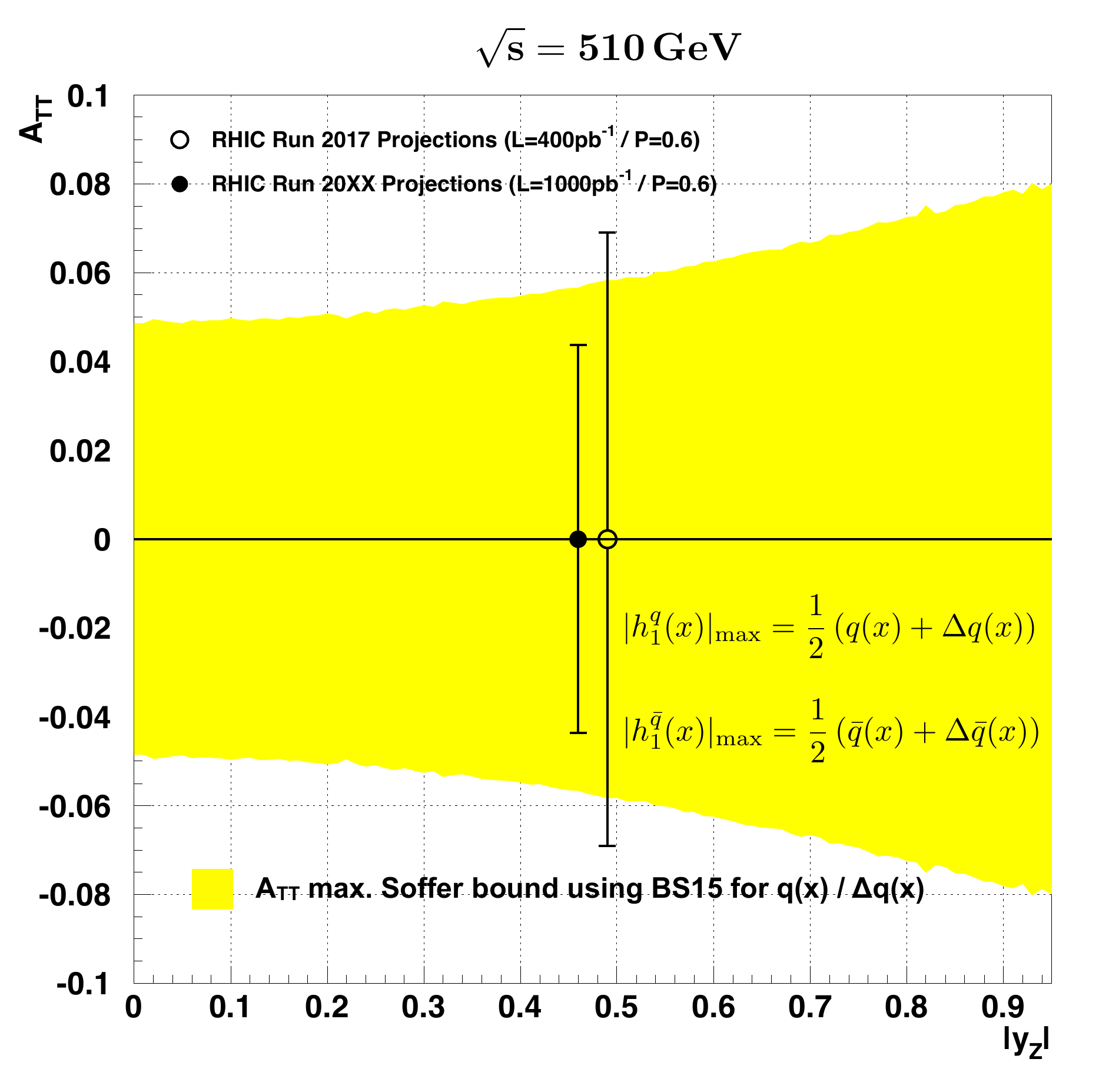}
\end{center}
\caption{\small Maximal bound of $A_{TT}$ as a function of $|y_{Z}|$ shown as a shaded band in comparison to uncertainty estimates for two running 
scenarios at RHIC in 2017 of $400\,$pb$^{-1}$ and a potential long run beyond 2020 of $1000\,$pb$^{-1}$ assuming a beam polarization of $60\%$.}
\label{fig4}
\end{figure}

The measurement of $A_{TT}^{Z/\gamma^{*}}$ is expected to be performed using the same selection criteria of $|\eta|<1$, $E_{T}>15\,$GeV and an 
invariant $e^{+}e^{-}$ mass window cut of $70\,<m_{\rm inv}\,<\,110\,$GeV. These cuts have therefore been used for a full NLO calculation. The unpolarized cross-section has been determined using the PDFs set of BS15 \cite{Bourrely:2015kla}. Figure~\ref{fig4} displays the maximal bound as a function $|y_{Z}|$ shown as a shaded band for the transversity distributions used in the calculation of $A_{TT}$. Note that $A_{TT}(y_{Z})=A_{TT}(-y_{Z})$. The respective LO result is very similar suggesting that NLO corrections are in fact rather small. The maximal bound is compared to two running scenarios at RHIC in 2017 of $400\,$pb$^{-1}$ and a potential long run beyond 2020 of $1000\,$pb$^{-1}$ assuming a beam polarization of $60\%$. The uncertainties on $A_{TT}$ for $Z$ production have been evaluated using the projected 2017 and beyond 2020 luminosities, beam polarization values at RHIC together with published experimental efficiencies for $Z$ production \cite{STAR:2011aa}. Systematic uncertainties for double-spin asymmetries are expected to be 
much smaller \cite{Adamczyk:2014ozi, Surrow:2015iqa}.

A measurement of a $A_{TT}$ for $Z$ production would provide the first direct constrain of transversity distributions. The sensitivity for $A_{TT}$ for W boson production has been also estimated. The uncertainties of $A_{TT}$ for $W^{+}$ / $W^{-}$ boson production at $400\,$pb$^{-1}$ ($1000\,$pb$^{-1}$), amount to $0.015$ / $0.024$ ($0.009$ / $0.015$), assuming also a beam polarization of $60\%$, taking into account published experimental efficiencies for $W^{+}$ / $W^{-}$ boson production \cite{STAR:2011aa}.
The observation of a statistically significant non-zero $A_{TT}$ for $W$ production would violate expectations within the Standard Model.
\clearpage
\section{Summary}
We have emphasized the relevance of measuring the transverse double-spin asymmetries $A_{TT}$ in the production, at mid-rapidity, of $W$ and $Z/\gamma^*$ gauge bosons in polarized proton-proton collisions at a center-of-mass energy $\sqrt{s} = 510 \mbox{GeV}$.\\
For $W$ production $A_{TT}=0$ beyond LO is a very strong result. It will be checked for the first time in the 2017 run at RHIC at BNL. We have also presented a QCD NLO calculation of the allowed maximal $A_{TT}$ for $Z$ production, by saturating Eq. (\ref{Delta}) to construct the maximal bound for quark (antiquark) transversity distributions $h_{1}^{q}(x,Q^2)$ $(q=u,d)$, from the PDFs set of BS15 \cite{Bourrely:2015kla}. Although the experimental uncertainty in this case is larger, due to a much lower production rate and the limited acceptance of STAR, this will be the first attempt to extract directly transversity distributions from hadronic collisions, in particular for antiquarks barely unknown. The situation is better with long-term prospects beyond 2020 at RHIC.\\
Finally, in addition to the Drell-Yan production mechanism involving only transversity distributions, so far considered, there could be other terms involving the product of TMD parton distributions, called double Sivers effect or double worm-gear effect \cite{Huang:2015vpy}. They require a non-zero low transverse momentum for the gauge boson and the corresponding transverse double-spin asymmetries are expected to be very small.

\begin{center}
{\bf Appendix}
\end{center}
Here we present a simple proof that $A_{TT}=0$ beyond LO, for $W$ production. At LO in the Drell-Yan picture one has the quark-antiquark fusion reactions $q(p_1,s_1) + \bar q(p_2,s_2) \to W$, where $p_1,p_2$ are the four-momenta of $q$ and $\bar q$, $s_1,s_2$ their corresponding spin pseudo-vectors. The matrix element describing the above transition, assuming a universal $V - A$ coupling, is $M = \bar u_q(p_1,s_1)\gamma_{\mu}(1 - \gamma_5)u_{\bar q}(p_2,s_2)W^{\mu}$, where $u(p,s)$ is the four-components spinor $u(p,s)$. The projection operator onto a state of definite spin $s$ is $u(p,s) \bar {u}(p,s)=1/2(p\!\!\!/+ m)(1 + \gamma_5 s\!\!\!/)$, using the Feynman slash notation defined as $k\!\!\!/ = \gamma^\mu k_\mu$. The factor $(1 - \gamma_5)$ in $M$ stands for the fact that the $W$ are left-handed. 
For the transverse double-spin asymmetry $A_{TT}$, one needs to take a transverse spin vector $s_T$ and, in the massless case, the cross section  $M M^ \dagger$ involves $Tr[p\!\!\!/_1(1 + \gamma_{5}s\!\!\!/_{T1})\gamma_{\mu}(1 - \gamma_5)p\!\!\!/_2(1 + \gamma_{5}s\!\!\!/_{T2})\gamma_{\nu}(1 - \gamma_5)]$. If one moves to the left, the most right $(1 - \gamma_5)$ factor, it reaches the other $(1 - \gamma_5)$ factor after three sign changes and becomes $(1 + \gamma_5)$, so it is obvious that this trace is zero. Now at NLO one can start radiating a gluon. The gluon is a vector particle, so one adds another $\gamma_{\alpha}$ to the trace and also a quark propagator $p\!\!\!/$. The sign changes survives, the trace still vanishes and this is true for as many gluons as one wants. In the case of the double helicity asymmetry $A_{LL}$, one uses the projection operator onto a state of definite helicity, $u(p,h) \bar {u}(p,h)=1/2(p\!\!\!/+ m)(1 + \gamma_{5} h)$, so clearly one does not get zero.

\vspace*{+5mm}

{\bf Acknowledgments}\\
We are grateful to W. Vogelsang for clarifying several points of this work and for providing a useful code to make
some calculations. We also express our sincere gratitude to D. de Florian for providing help with the unpolarized NLO Z production code. We greatly appreciate 
interest and encouragement from A. Metz at an earlier stage of this work.

\end{document}